\documentclass[twocolumn,prl,aps,floatfix,superscriptaddress]{revtex4}

\usepackage{graphicx}
\usepackage[all]{xy}

\usepackage{amsthm}
\usepackage{amsfonts}

\newcommand{\be}{\begin{equation}}
\newcommand{\ee}{\end{equation}}

\newcommand{\ct}{{\rm const.} \times}
\newcommand{\rinj}{R_{inj}}
\newcommand{\rdec}{R_{dec}}

\newcommand{\cred}{c_{red}}
\newcommand{\pc}{T_{poor}}

\newtheorem*{theorem*}{Theorem}
\newcommand{\Hfour}{\mathbb{H}^4}

\begin{document}
\title{Decoding in Hyperbolic Spaces: LDPC Codes With Linear Rate and Efficient Error Correction}

\author{Matthew B. Hastings}
\affiliation{Microsoft Research, Station Q, CNSI Building, University of California, Santa Barbara, CA, 93106}

\begin{abstract} 
We analyze the four dimensional toric code in a hyperbolic space and show that it has a classical error correction procedure which runs in almost linear time and can be parallelized to almost constant time, giving an example of a
quantum LDPC code with linear rate and efficient error correction.
\end{abstract}
\maketitle

Many quantum codes that are studied are based on the stabilizer formalism.  Of these, the low-density parity check (LDPC) codes are particularly interesting.  A quantum LDPC code, for our purposes, is a quantum stablizer code in which all of the stabilizers act on at most $O(1)$ qubits and in which each qubit participates in at most $O(1)$ stabilizers.  Particular examples of quantum LDPC codes include the toric code\cite{tc} and various generalizations of it discussed below, as well as hypergraph product codes\cite{hyperproduct,kp}.

Quantum LDPC codes with a linear rate have been invented.  Perhaps the earliest example of such was the two-dimensional toric code on a surface of constant negative curvature\cite{fm}, giving an example with a logatihmic distance.  Later examples include the hypergraph product codes which obtain $\Theta(\sqrt{N})$ distance, where $N$ is the number of qubits.  Finally, in Ref.~\onlinecite{hyperfinite}, the four-dimensional toric code\cite{4dtoric} was considered in a four-dimensional hyperbolic space was considered, and again shown to have linear rate.  Ref.~\onlinecite{guthlubotzky} further studied this four-dimensional code and provided explicit constructions of the needed hyperbolic four-manifolds, showing one could attain a logarithmic injectivity radius for these manifolds and hence a distance that scaled proportional to $N^\alpha$ for some $0< \alpha<1$.
The injectivity radius $\rinj$ of a hyperbolic manifold is defined to be at least $R$ if, for any point $x$ in the manifold, the ball of radius $R$ around that point is isometric to the ball of radius $R$ in hyperbolic space $\Hfour$.  Note that there are no nontrivial closed geodesics or nontrivial minimal surfaces in a ball of diameter less than the injectivity radius; intuitively, this fact is why the injectivity radius is relevant for the distance of the code.  

In this paper, we further consider this four-dimensional code and show that it has a threshold and an efficient classical decoding procedure; each round of the decoding can be parallelized to run in constant time and correcting to a codeword takes logarithmically many rounds.
The classical procedure is a greedy local procedure that ``shrinks" error chains of length slightly smaller than the injectivity radius.

One reason for the interest in these properties (quantum LDPC code with linear rate and efficient classical decoding) is that recently\cite{daniel} it was shown that such a code would allow one to perform fault-tolerant quantum computation with only a constant factor overhead.  The construction of Ref.~\onlinecite{guthlubotzky} gives a sequence of manifolds $M_j$ for such that $N$ increases polynomially in $j$; hence, 
these codes meet condition (iii) of the main result of Ref.~\onlinecite{daniel} concerning how frequent the members of the code family are.
The bounds below give an error probability that is bounded by $\tau \times (\ct p)^{\ct \log(N)}$, where $\tau$ is the time for which the computation is run.  Hence, for any fixed polynomial scaling of $\tau$ with $N$, there is an error threshold.  Given that these codes have a distance that scales as $N^\alpha$, it is possible that the error probability in fact scales to zero more rapidly with $N$ than this estimate; if so, then there would be a single error threshhold for all polynomial $\tau$.

We analyze the code in the context of its use as a quantum memory, rather than just as a quantum channel.  That is, rather than considering a model in which information is perfectly encoded into some code state, then noise is applied, then finally one attempts to decode using perfect quantum gates, we assume instead that after the information is initially encoded, the information much be maintained for many time steps.  On each time step, some noise is applied followed by some (possibly imperfect) measurements and some (also possibly imperfect) error correction is applied.  This protocol is discussed further below, as is the noise model which is the same adversarial noise model as used in Ref.~\onlinecite{daniel}.  In this model, the adversary is not allowed to select the errors in a completely arbitrary fashion; however, one also does not assume complete independence of different errors, instead assuming that the probability of having errors on any given set $X$ is bounded by $p^{|X|}$ for some $p>0$.

Before giving any formal details, let us give a purely heuristic motivation for why such an error correction procedure might work: the error syndrome in this code consists of several closed one dimensional loops.  In $\mathbb{R}^4$, a closed loop of large radius may have a small amount of curvature locality, with that amount of curvature going to zero as the loop becomes large.  However, in $\Hfour$, because of the negative curvature, even a large closed loop must have large curvature somewhere.  This allows a greedy procedure in which we try to shorten the loops locally.  We actually take advantage of two different ways of shortening a loop.  One way is to shorten a loop while leaving it as a single loop.  Another way is to split it into two or more smaller loops.
For an example of this, consider a closed loop in $\Hfour$ which is a geodesic triangle.  In this case, every point on any given side of the triangle is within some bounded distance of one of the other two sides (this is because it is a so-called $\delta$-hyperbolic space); this allows us to perform local moves in which we split the triangle into two smaller loops.

\section{Toric Codes in Hyperbolic Spaces}
We begin by reviewing the toric code and its generalizations, sometimes called ``homology codes".
In general, given a manifold and a cellulation of that manifold, one can define a toric code.  In the two dimensional toric codes, the degrees of freedom are associated with the $1$-cells while the $Z$ stabilizers are associated with the $0$-cells and the $X$ stabilizers are associated with the $2$-cells.  Each $Z$ stabilizer acts on the qubits in its coboundary while each $X$ stabilizer acts on the qubits in its boundary.  The commutativity of the stabilizers is guaranteed by the fact that the ``boundary of a boundary is zero".
In Ref.~\onlinecite{fm}, it was shown that a two dimensional toric code could have a linear rate.  There, the code was considered on a cellulation of a family of surfaces of constant negative curvature.  Fixing the curvature to $-1$ for all of these surfaces, the $2$-cells were all taken to have a volume of order unity, so that the total number of cells was proportional to the volume of the manifold.  Hence, the number of qubits scaled with the volume of the manifold.
Because of the constant negative curvature, by the Gauss-Bonet theorem the genus of the surface was proportional to the volume, giving the linear rate (since the qubits are associated with the $1$-cells, the number of logical qubits is equal to the first Betti number using $Z_2$ homology).
The injectivity radius in this family grew logarithmically with the volume of the manifold, giving a logarithmic distance.

 It is possible that that family of codes could meet the requirements of having an efficient classical decoding algorithm, even against the kind of adversarial noise considered in Ref.~\onlinecite{daniel}.  A likely candidate for the decoding algorithm would be minimum weight perfect matching.  However, in this paper we consider a slightly different family of codes for which the analysis of the decoding algorithm is simpler; for this other family, a simpler greedy local classical decoding algorithm suffices.

We consider a family of four-dimensional manifolds.  In this case, we use a four-dimensional toric code\cite{4dtoric} so that the degrees of freedom are associated with the $2$-cells
while the $Z$ stabilizers are associated with the $1$-cells and the $X$ stabilizers are associated with the $3$-cells.  The number of logical qubits is equal to the second Betti number using $Z_2$ homology.
We consider a family of four-dimensional manifolds with constant negative curvature (fixed to $-1$) and diverging injectivity radius.  These codes based on these manifolds were first discussed in Ref.~\onlinecite{hyperfinite} where they were shown to have linear rate.  In Ref.~\onlinecite{guthlubotzky} it was shown that the injectivity radius could be taken to diverge logarithmically with $n$, which will be essential below.

We triangulate this manifold with simplices to define the code.  Call this cell complex $K$.
We again choose to take all top cells to have a volume of order unity, so that the volume of the manifold is proportional to the number of encoded qubits.  Further, we can choose to take a bounded local geometry by the following theorem of Ref.~\onlinecite{bowditch} so that each $1$-cell will have length within a constant factor of unity.
\begin{theorem*}
Given an integer $d\geq 2$ and a real number $r>0$, there is constant $C(d,r)>0$ so that every hyperbolic $d$-manifold with injectivity radius $>r$ can be triangulated with geodesic $d$-simplicies $\sigma_i$ of bounded geometry in the sense that 
each $\sigma_i$ admits a homeomorphism $h_i:\sigma_i \rightarrow \sigma_0$, $\sigma_0$ the hyperbolic simplex with all sides of length $1$ so that
$$
\frac{1}{C(d,r)} d(x,y) \leq d(h_i(x),h_i(y) \leq C(d,r) d(x,y)
$$
for all $x,y\in \sigma_i$.
\begin{proof}
See Ref.~\onlinecite{bowditch}.
\end{proof}
\end{theorem*}

\section{Greedy Decoding}
We now describe a simple local greedy decoding algorithm.  We begin by describing the decoder and then analyze its performance assuming that no noise occurs during decoding.  In the next section, we study the application to quantum memories.

To be a good quantum code, we must show error correction against both $S^z$ errors (dephasing) and $S^x$ errors on qubits (spin flips).
In this section and the next, we analyze only the performance against spin flip noise.  This noise causes errors in the $Z$-stabilizers associated with the $1$-cells.  The analysis of the performance against dephasing noise can be analyzed in exactly the same way by working on a dual cellulation, so that the $X$-stabilizers are associated with $1$-cells.

We use the language of chain complexes.  A $k$-chain is a vector in a vector space whose basis elements correspond to the $k$-cells.
Since we work with $Z_2$ homology, the coefficients of this vector are either equal to $0$ or $1$.  When we refer to a cell being in the support of a chain, we mean that the coefficient is equal to $1$.  For brevity, we sometimes say that a $1$-cell is ``in" a chain, to mean that it is in the support.
If $C_k$ is a $k$-chain, we write $|C_k|$ to denote the number of $1$-cells in the support of $C_k$.  We write $\partial$ for the boundary operator.   

In a code state, all of the $Z$-stabilizers have expectation value $+1$.  After errors on some set of spins $D$, the $Z$-stabilizers in the support of $\partial D$ have errors.
Note that $\partial D$ is closed, so the set of errors always form a closed chain.  We call this the error chain (sometimes it is called the ``error syndrome").

We say that a closed chain $C$ is {\it atomic} if it cannot be written as the sum of two closed chains, both disjoint and both nonzero.
Any error chain $C$ can be written as a sum $C=\sum_i C_i$ where the $C_i$ are atomic closed chains.
We say that an atomic chain is {\it small} if the radius of its support is bounded by $\rinj-2 \rdec$, where the constant $\rdec$ is given below.
We say that an error chain $C$ is {\it small} if it can be written as a sum of small atomic chains.

\subsection{Greedy Decoder}
To define the decoder, we choose some $\rdec>0$. 
In the next subsection we will describe how to choose $\rdec$.  The choice of $\rdec$ will be independent of $N$ so for sufficiently large $N$, the injectivity radius will be much larger than $\rdec$ (in later sections we will simply treat $\rdec$ as a constant $O(1)$).  From now on we assume that we indeed are in the case that the injectivity radius is large compared to $\rdec$.

The decoder operates in several rounds.  In each round, we pick a set $X$ of random points in the manifold, each with distance at least $2\rdec$ from each other.
We then define a set of balls; each ball will be the set of points within distance $\rdec$ of one of these random points.  We pick these random points
such that every point has some strictly positive probability of being within distance $\rdec/2$ of one of the random points.
To do this, we pick a random set of points $Y$ indepdently with some fixed density $\rho>0$ and then let $X$ be the set
of points in $Y$ which are not within distance $2\rdec$ of another point in $Y$.  Then, the probabilities that several points $x_1,...,x_n$ are in balls are independent for sufficiently large separation between pairs $x_i,x_j$.
The centers of the balls will be located in the ambient space (the hyperbolic manifold) and need not be at $0$-cells of the triangulation.  We choose the centers generically so that no $0$-cell lies exactly distance $\rdec$ from the center of a ball; this is done to simplify the explanation of the algorithm below.

Next, we measure the stabilizers in each ball.
Then, we perform a greedy reduction in weight in each ball.
as follows.  For each ball, there are some $1$-cells that intersect the boundary of the ball which we call ``fixed" $1$-cells.  There are also $1$-cells which are contained in the interior of the ball, which we call ``variable" $1$-cells.  We do not consider the $1$-cells that are entirely in the exterior of the ball.  Then, for any given errors on the fixed $1$-cells, we find a minimum weight error chain on the variable $1$-cells that gives an error chain that is a closed $1$-chain in that ball; we emphasize that we choose ``a" minimum weight error chain rather than ``the" minimum weight error chain as there may be more than one.  If the original chain is minimum weight, then no spin flips are applied.  Otherwise, 
we apply a set of spin flips contained within the ball to produce this minimum weight error chain.  This set of spin flips can be computed as follows.  Let $C$ be the previous error chain (including all $1$-cells regardless of their position relative to the ball), and let $C'$ be the error chain which results from changing $C$ on the variable $1$-cells to minimize the weight.  Then, $C$ and $C'$ are homologous, with $C=C'+\partial D$ for some $2$-chain $D$ contained in the ball.  Finally having calculated $D$ we apply
 spin flips on the $2$-cells in the support of $D$.  Note that the error-correcting process can be done in parallel on different balls.

In the next subsection we analyze this decoder.  In the subsection after this, we give a modification to a deterministic decoder.

\subsection{Analysis of Decoder} 
We now show that for sufficiently large $\rdec$, if the error chain $C$ is small, then the greedy decoder will reduce the number of errors by at least a constant fraction in each round on average.
Further, we show that the probability that the weight is not reduced by a factor $\cred$, for some $\cred<1$, is bounded by $\exp(-\ct |C|)$, for some positive constant.

The calculation in this section assumes perfect measurement of the error chain by the decoder and assumes that no noise is applied in between rounds or during the correction process.  In the next section we consider the effects of imperfect measurements and also noise over many rounds of application of the decoder.

We consider first the case that $C$ is an atomic closed chain. 
We analyze the decoder in three steps.  First, we construct a certain planar complex whose boundary maps to the boundary of $C$ and which has certain other nice properties described below, including an bound on the area of the complex from an isoperimetric inequality.  Second, we use a planar separator theorem to decompose the complex into sets with small boundary between the sets.  Third, we analyze the decoder by analyzing the effect of flipping the spins in the image of any one of these sets.

\subsubsection{Construction of Complex}
We claim that $C=\partial P$ for some chain $P$ such that $|P|\leq \ct |C|$.  Further, we can
identify a $2$-complex $M$ and a continuous mapping $f$ from $M$ to the $2$-skeleton of $K$ such that the following properties hold.
First, every $2$-cell of $M$ is mapped onto one $2$-cell of $K$.  Second, $M$ is planar; in fact, $M$ is a disk, and the boundary of $M$ is mapped one-to-one to $C$ and hence the chain which is the sum of all $2$-cells in $M$ is mapped to a chain
$P$ such that $C=\partial P$.
Third, the number of cells of $M$ is at most $\ct |C|$.  Fourth, every $2$-cell of $M$ is attached to $O(1)$ $1$-cells.
It is possible that the mapping from $M$ to $K$ is many-to-one in the interior, and that several $2$-cells in $M$ may map to the same $2$-cell in $K$. 
The construction of this chain $P$ is by constructing a discretization of a minimal spanning surface; the purpose of introducing $M$ is to allow us to parametrize this surface in the case that it is not an embedded surface (for example, it may have intersections).

To construct $M$,
given an atomic error chain, we define a closed curve in the ambient space in the natural way: each $1$-cell in the error chain defines some arc of length at least $1/C(d,r)$ and at most $C(d,r)$ and the curve is simply the union of the arcs in the $1$-cells in the chain.
Call this curve $\gamma$.
Let $\Sigma$ be the minimal surface whose boundary is $\gamma$, choosing $\Sigma$ to be the image of a disk $D$ under a continuous map with the boundary of $D$ mapping one-to-one to $\gamma$ (we allow the surface to be an immersed disk rather than embedded).  By the Gauss equation, since $\Sigma$ is minimal and since the ambient space has constant curvature equal to $-1$, $\Sigma$
is a surface whose curvature is bounded above by $-1$.  So, by an isoperimetric inequality we can bound the area of $\Sigma$ by a constant times the length of $\gamma$.  The surface $S$ can be deformed to lie entirely on the $2$-skeleton by moving each point on the surface at most a bounded distance and increasing the area by at most a constant factor.
To do this deformation, first choose a random point in each $4$-simplex, and cone outwards from that point so that $\Sigma$ lies entirely on the $3$-skeleton (i.e., map each point in $\Sigma$ to the point on the boundary of the $4$-simplex which lies on a geodesic from the given random point through the given point on $\Sigma$).  Then, choose a random point in each $3$-simplex and again cone outwards so that $\Sigma$ lies entirely on the $2$-skeleton.
Choosing these points at random, the average area increase is bounded.
Let $P$ be the $2$-chain corresponding to that deformed surface (i.e., $P$ is the sum over $2$-cells in the deformed surface, with appropriate multiplicity), so that $C=\partial P$ and $|P| \leq \ct |C|$ as claimed.
To define $M$, pick any point $x$ in the disk $D$ which does not lie in the $1$-skeleton so that the image of $x$ lies in the interior of some $2$-cell $\sigma$; define a $2$-cell around this point consisting of all points in the disk that can be reached by a path entirely in $\sigma$.  Repeat this for different points in the disk until the disk is covered with cells.  If the image of any of these cells does not cover the corresponding $2$-cell $\sigma$, then we can deform the map to remove this $2$-cell (i.e., by coning outwards from any missing point in $\Sigma$).  
Then since each remaining cell has an area at least equal to that of the smallest $2$-cell in $K$ (as otherwise we could remove it), the number of these $2$-cellls in $M$ is at most $\ct |C|$ as claimed.
Since $D$ is a disk, we can embed the $2$-complex $M$ above in the plane.

This construction may not yet give the property that every $2$-cell of $M$ be attached to a bounded number of $1$-cells.  We will use this property below to ensure that a certain graph has bounded degree.  Consider some $2$-cell $\tau$ in $M$ which is mapped to some $2$-cell $\sigma$ in $K$.  In general, $\sigma$ is attached to $O(1)$ different $1$-cells.  Hence, 
if $\tau$ is attached to some large number of $1$-cells, then the map on the image of the boundary of $\tau$ must not be one-to-one.  For example, if $\sigma$ neighbors three different $1$-cells denoted $\tau_A,\tau_B,\tau_C$ in $K$, then it is possible that the boundary of $\tau$ maps to $\tau_A,\tau_B,\tau_C,\tau_A,\tau_B,\tau_C,\tau_A,\tau_B,\tau_C,...$ in sequence.   It is not clear if such a situation can arise from the construction above; for example, given a bound on the mass of $\Sigma$ in all $4$-simplices (which seems very plausible), then (since mass increases by only a constant amount under the deformation) this situation would be forbidden and we would have the desired bound on the number of $1$-cells attached to a $2$-cell.  However, we can also address this situation in a purely combinatoric fashion without referring to $\Sigma$.
If such a case occurs, 
since the map on the boundary of the cell is not one-to-one we can split $\tau$ into two $2$-cells by deforming the map to be constant on some path in the $2$-cell which connects two boundary points with the same image and then identifying points on that path.  We choose these points to have their image on $0$-cells of $K$ (for example, the $0$-cell attached to $\tau_A$ and $\tau_B$ would be such a point).  Then, by doing this identification we split $\tau$ into smaller $2$-cells.
Indeed, if the degree of the map of the boundary of $\tau$ is larger than one, we can split $\tau$ into cells whose boundary map has smaller degree.  We proceed with this process until for any given $2$-cell $\tau$ of $M$, the boundary of $\tau$ maps to a sequence of $1$-cells in $K$, with each $1$-cell appearing only once.  This step gives the desired bounded degree property.  Note that if after this step,
if the image of any of $2$-cell $\tau$ in $M$ does not cover the corresponding $2$-cell $\sigma$ in $K$, then we can again deform the map to remove this $2$-cell, so we retain the property that the number of these $2$-cellls in $M$ is at most $\ct |C|$.
This completes the description of the construction of the complex.

In the above construction, the fact that the chain is small is used to show that, having radius smaller than $\rinj-2\rdec < \rinj$, we can map a ball containing that chain isometrically to a ball in $\Hfour$.  Then, we can find the minimal curvature surface $\Sigma$ in $\Hfour$.  We can assume that $\Sigma$ is still contained inside the ball of radius $\rinj-2\rdec$ as otherwise it would not be minimal, and so we can map $\Sigma$ back to the compact hyperbolic manifold isometrically and then deform it to lie on the $2$-skeleton after mapping back.

\subsubsection{Planar Separator Theorem}
We now claim that for any $m$, the sum of all $2$-cells in $M$ can be decomposed as a sum of disjoint chains $M_1,M_2,...$ each containing at most $m$ cells such that the total number of $1$-cells which are in the boundary of both $M_a$ and $M_b$ for some pair $a\neq b$ is bounded by $O(|C|/\sqrt{m})$; equivalently, the total
number of $1$-cells which are in the boundary of some $M_a$ but not in the boundary of $M$ is at most $O(|C|/\sqrt{m})$.  
The existence of these sets follows from a version of the planar separator theorem\cite{pst1}.  We use the version in Ref.~\onlinecite{pst2} which shows that given any planar graph with $V$ vertices, for any $\epsilon$, we can remove at most $4 \sqrt{V/\epsilon}$ vertices to give a new graph with no component having more than $\epsilon V$ vertices.  The planar graph we consider is a graph whose vertices correspond to $2$-cells in $M$, with an edge between two different $2$-cells if they are attached to each other in $M$.  We pick $\epsilon=\ct m/V \geq \ct m/|C|$ so that each component has at most $\ct m$ vertices in it.  There are then $O(V/\sqrt{m})=O(|C|/\sqrt{m})$ removed vertices.  We then take each removed vertex and add it to one of the components that neighbor that removed vertex; this increases the size of the components by at most a constant factor so that each still has at most $m$ vertices (here, the bounded degree of the graph is used to show that the number of cells in the boundary of any given component is also increased by only a constant factor).  Thus, rather than constructing a separator that removes vertices, we construct a separator that removes edges.  Let the resulting components be denoted by sets $V_1,V_2,...$ 
Finally, we identify the chains $M_a$ with the different components; more precisely, for each component $V_a$, $M_a$ is the sum of $2$-cells in $V_a$.
This completes the construction of the chains $M_a$.

\subsubsection{Greedy Decoder and Flipping Spins in $f(M_a)$}
We now use the chains $M_a$ to analyze the performance of the greedy decoder, and choose $\rdec$.
Let
\be
P_a=f(M_a),
\ee
so that
\be
P=\sum_a P_a
\ee
and
$C+\sum_{a} \partial P_a=0$. 
We can assume that the $M_a$ are connected, treating two $2$-cells as connected if they are attached to each other (the construction above makes them connected, but even if they were not, we can simply split them into connected components).
The basic idea is that since each $V_a$ has only a small boundary, doing the spin flips corresponding to cells in $P_a$ will tend to reduce the weight of $C$ (doing these spin flips may help by cancelling cells in $C$ but may hurt by creating other cells in the boundary of $P_a$).
 Define $N^{bulk}_a$ to be the number of $1$-cells in
$\partial M_a$ which are not in the boundary of $M$.  Define $N^{bdry}_a$ to be the number of $1$-cells in $\partial M_a$ which are in the boundary of $M$.
Define $\Delta_a=N^{bulk}_a-N^{bdry}_a$.
Then
given any set $X$ whose elements are chosen from the set of possible indices $a$, we have
\be
\label{partsum}
|C+\sum_{a \in X} \partial P_a| -|C| \leq \sum_a \Delta_a.
\ee
Also,
\be
|C|+\sum_{a} \Delta_a- \sum_{a} N^{bulk}_a=0.
\ee
However, by construction
$\sum_{a} N^{bulk}_a \leq O(|C|/\sqrt{m})$ so
\be
\sum_{a} \Delta_a\leq -|C| (1-O(1/\sqrt{m})).
\ee
Choosing $m$ sufficiently large that  $(1-O(1/\sqrt{m})) \geq 1/2$, we have
\be
\sum_{a} \Delta_a\leq -(1/2) |C|.
\ee

Since each chain $P_a$ is connected and has at most $m$ cells, it has bounded diameter.  We
choose a sufficiently large $\rdec$ so that each set $P_a$ has diameter at most $\rdec/2$.  Given this choice, each $P_a$ has a strictly positive probability of
being in a ball given the random choice of balls; this probability depends upon the density of the balls.
Hence, by Eq.~(\ref{partsum}) by performing the spin flips corresponding to all the $P_a$ which fall in a ball in the given round, the decoder on average reduces the weight of $|C|$ by a constant fraction.  The greedy decoder will always perform at least this well (it may find a way of even further reducing the weight) and hence also on average reduces the weight by a contant factor.
Further, for  sets $P_a$ which are far from each other, the probabilities that they are in a ball are independent and so
the probability that a constant fraction of them are not in a ball is exponentially small in $|C|$.

Hence, with probability $1-\exp(-\ct |C|)$, the chain after error correction has weight at most $\cred |C|$, for some $0\leq \cred<1$.
The above calculation was for an atomic chain $C$.  However, given a non-atomic chain $C$, we write it as a sum of atomic chains and apply the same construction to each atomic chain (given sets $P_a$ for each atomic chain, the greedy decoder does at least as well as it would be applying the spin flips in those sets).  Hence the same weight reduction holds for a non-atomic chain. 

\subsection{Deterministic Correction Scheme}
The decoder above was randomized.  This has some advantage that little global coordination is required; one can imagine an implementation where local classical processing elements independently try at random times to perform a local error correction scheme.  Each classical processing element could try to correct within a given ball; it would communicate locally to see if the ball it is trying to correct overlaps with any nearby classical element; if not, it would try to correct and if yes it would wait.

However, we might prefer a deterministic scheme.  To do this, cover the space with overlapping balls of radius $\rdec$ so that every point is within distance $\rdec/2$ of the center of one of the balls.  Color the balls with $k=O(1)$ colors such that no two balls of the same color are within distance $2\rdec$ of each other.  Then, a single decoding round in the deterministic scheme is broken into $k$ subrounds.  In each subround, we perform a greedy decoding process in the balls of the given color.
Following the above analysis, we know that given a chain with given weight $|C|$, there are at least $\ct |C|$ different balls in which the greedy decoding process can reduce the weight by at least $1$.  Now, if a given ball $B$ overlaps with some other ball $B'$, and if decoding is performed on $B'$ before it is performed on $B$, then even if ball $B$ could have reduced the weight by at least $1$ if it were performed first, ball $B'$ might modify the chain in such a way that it is no longer possible for $B$ to reduce the weight.  However, since each ball overlaps with only $O(1)$ other balls, and since there are only $O(1)$ other rounds, there are only $O(1)$ different balls $B'$ whose error corrections is done before $B$ such that the error corrections done on $B'$ will
 change the state on $B$.  Further, since error correction is only done on $B'$ if the weight can be reduced by at least $1$, one can see that the deterministic scheme still leads to at least a constant reduction in weight after every $k$ subrounds.
To do this estimate more formally, let $W_a$ be the weight that would be reduced in that $a$-th round (for $a=1,...,k$) assuming that no error corrections was done on any previous rounds.  Let $W'_a$ be the actual weight that is reduced on the $a$-th round given the result of the previous rounds.  Note that $W'_a \geq W_a - \ct \sum_{b < a} W_b$ using the fact that each ball only overlaps with $O(1)$ other balls. 
Hence, $\sum_a W'_a \geq \ct \sum_a W_a$.

\section{Quantum Memory}
We now consider this code as a quantum memory.  We use a discrete time model.  In each time step,  some syndrome measurements are performed, with errors possibly occuring in these measurements.  Then, the correction procedure is applied to certain qubits, again with error possibly occuring when the corrections are applied.  Finally, additional errors may be applied to all qubits.  In each time step, the syndrome measurements that we perform will be those needed to perform a given round of the correction procedure above.

We analyze the randomized decoder above.  The analysis of the deterministic scheme can be done in the same way.

The error model considered is that on any given time step, given any set $X$ of qubits and any set $S$ of stabilizers being measured, the probability of having errors on those qubits and those stabilizer measurements is bounded by $p^{|X|+|S|}$ for some $0<p<1$.  Note that this is distinct from having errors occuring in an independent fashion with probability $p$ of an error and probability $1-p$ of no error; all we do is bound the probability of having errors without assuming independence.

In practice, each stabilizer measurement would be performed by some quantum circuit consisting of CNOT and Hadamard gates, followed by a single measurement on some ancilla qubit; since each stabilizer has weight $O(1)$, there are $O(1)$ possible places for errors to occur in this quantum circuit.  Thus, we could also consider a quantum circuit model in which the probability of having errors in any set $G$ of gates in the circuit (including the final measurement and including identity gates for qubits on which we do not act) is bounded by $q^{|G|}$ for some $q>0$.  Since the number of gates $|G|$ needed to measure $|S|$ distinct stabilizers is at most a constant factor larger than $|S|$, this quantum circuit error model can be fit into the error model described above by taking $p=q^{O(1)}$.  One might worry that we might need to measure many overlapping stabilizers and hence it would not be possible to start measuring a given stabilizer $Z$ stabilizer until one has finished measuring any $X$ stabilizers with which that $Z$ stabilizer overlaps and that this might increase the depth of the quantum circuit.  However, this leads to an increase in depth by a multiplicative factor that is $O(1)$ since each stabilizer overlaps with only $O(1)$ other stabilizers and so again by taking $p=q^{O(1)}$ we can fit it into the error model above.

Using this error model, given any set of time slices, $1,...,\tau$, the probability of having errors on sets of qubits $X_1,...,X_\tau$ and on sets of stabilizers $S_1,...,S_\tau$ in the corresponding time slices is bounded by
\be
\label{PerrTot}
P_{error}\leq p^{\sum_{i=1}^\tau (|X_i|+|S_i|)}.
\ee

We describe the process of spin flips by a diagram in spacetime.  Let $K$ be the cell complex that is the cellulation of the four-manifold used to define the code.  Let $I$ be a cell complex that is a cellulation of an interval, with $\tau+1$ $0$-cells and $\tau$ $1$-cells, labelling the $0$-cells by $0,...,\tau$ and labelling the $1$-cells by $1,...,\tau$.  Let $K'$ be the product of $K$ with $I$.  Thus there are two types of $1$-cells in $K'$.  For every $1$-cell in $K$, there are $\tau+1$ different $1$-cells in $K'$ arising from the product of that given $1$-cell in $K$ with a $0$-cell in $I$.  Additionally, for every $0$-cell in $K$ there are $\tau$ different $1$-cells in $K'$ arising from the product of that given $0$-cell with a $1$-cell in $I$.
Similarly, for every $2$-cell in $K$, there are $\tau+1$ different $2$-cells in $K'$ arising from the product of that given $2$-cell in $K$ with a $0$-cell in $I$; we call these the spacelike $2$-cells and we write them $q \times i$ where $q$ labels a $2$-cell in $K$ and $0 \leq i \leq \tau$.  Additionally, for every $1$-cell in $K$ there are $\tau$ different $2$-cells in $K'$ arising from the product of that given $1$-cell with a $1$-cell in $I$; we called these the timelike $2$-cells and we write them $s \times i$ where $s$ is a $1$-cell in $K$ and $1 \leq i \leq \tau$ labels $1$-cells in $I$..
Given a set of spin flips, we define a $2$-chain $C$ as follows.  A spacelike $2$-cell $q \times i$ will be in $C$ if and only if there is a spin flip at the qubit corresponding to $q$ in timeslice $i$.  This spin flip can be due either to noise or to the recovery procedure.  A timelike $2$-cell $s \times i$ will be in $C$ if and only if the error chain after the $i-1$-th time slice includes the stabilizer corresponding to $s$.
Note that by construction, $C$ is a closed chain.

We write $C$ as a sum of closed chains $C=C_1+C_2+...$, as follows.  Consider a graph $G$ with vertices corresponding to $2$-cells and with an edge between vertices corresponding to two spacelike $2$-cells $q \times i$ and $r \times j$ if $i=j$ and if $q$ and $r$ are within distance $2\rdec$ of each other and with an edge between vertices corresponding to a spacelike and timelike $2$-cell or to two timelike $2$-cells if they are attached to each other.  We choose the $C_i$ to be the connected components of the support of $C$ given this graph.

What we will show is that for any $2$-cell $x$,
 the probability that $x$ is in a connected component $C_i$ which is not small is bounded by 
\be
\label{ebnd}
(cp)^{\rinj/c'}
\ee
for some constants $c,c'$.  Hence, for any $\alpha<\infty$, for sufficiently small $p$, the probability that any $C_i$ is not small is $O(N^{-\alpha})$.
Summing over $2$-cells $x$, the probability that there is any connected component $C_i$ which is not small is $O(\tau N^{1-\alpha})$.
After doing this, in subsection \ref{errcor} we will show that for any $\tau$ that is polynomially long in $N$, for sufficiently small $p$ this implies the ability to error correct the state back to a codeword with high probability.

From here on, for notational simplicity, we write $E$ to denote the $2$-chain that is the connected component of $C$ containing the fixed $2$-cell $x$.
We write $E(i)$ to indicate the set of errors before a given time slice.  More precisely, consider all $2$-cells in $E$ which are the product of a $1$-cell in $C$ and a $1$-cell in $I$ with the $1$-cell equal to $i$.  Let $E(i)$ denote the set of those $1$-cells in $C$ appearing in this product.

Let $w_i=|E(i)|$.
From the analysis before, with probability $1-\exp(-\ct w_i)$ we have a choice of balls on the $i$-th
round of error correction such that, without spin flip or syndrome errors, we would have $w_{i+1} \leq \cred w_{i}$ for some $0\leq \cred < 1$.  If a choice of balls is such that this would not happen without errors in the given round, we call this a {\it poor choice}.

We bound the probability of $x$ being in a connected component which is not small in two steps.  First, we show that for any $E$ with given $|E|$, a large number of errors (the number being proportional to $|E|$) must have occured or there must have been some number of poor choices.  We can upper bound the probability of having a given set of errors or poor choices; we then sum over possible sets of errors and poor choices and show that the probability of
having a given error chain $E$ is bounded by
\be
\label{PE}
P(E) \leq (\ct p)^{\ct |E_{spacelike}|}\cdot O(1)^{|E_{timelike}|},
\ee
where $E_{timelike}$ is the set of timelike $2$-cells of $E$ and $E_{spacelike}$ is the set of spacelike $2$-cells of $E$; the spacelike $2$-cells of $E$ arise from the error correction process as well as from spin flip errors.
Eq.~(\ref{PE}) is derived below.
Then, we bound the number of $E$ which include some given $2$-cell $x$ with given $|E_{spacelike}|$ and $|E_{timelike}|$ and apply a union bound by summing over all $E$, weighted by $P(E)$.

We start by showing Eq.~(\ref{PE}).
Note that by choosing the graph as we have done, all other connected components of $C$ are sufficiently far from $E$ that they do not affect the error correction process; i.e., no ball of radius $\rdec$ will include both $E$ and some other connected component.  This permits us to analyze the connected components separately from each other.

If $w_{i+1}>\cred w_{i}$ then there must be at least one error (either a stabilizer measurement or a spin flip error) on the $i$-th round
or there must be a poor choice.  Let $\pc$ be the set of rounds on which there is a poor choice.  The probability of
having poor choices on a given set of rounds $\pc$ is bound by $\exp(-\ct \sum_{i\in \pc} w_i)$.

By a triangle inequality $|E_{spacelike}|$ is bounded by $W^{ec}+W^{err}$, where $W^{ec}$ is the number of spin flips produced by the error correction assuming no errors in stabilizer measurements, and $W^{err}$ is the number of spin flips due to errors, either spin flip or stabilizer error measurements (i.e., $W^{err}$ is the weight of the difference between the
spin flips that would be applied without errors and the spin flips that are actually applied due to the combination of recovery and error).  
Further, $W^{err} \leq \ct N_{err}$, where $N_{err}$ is the number of errors (either qubit or stabilizer) within distance $\rdec$ of $E$; a single stabilizer error can cause at most $O(1)$ additional spin flips.

We now use an amortized analysis to show that $|E_{spacelike}| \leq \ct N_{err}$.
Let $W^{err}_i$ and $W^{ec}_i$ denote the number of spins flips due to errors and due to error correction on the $i$-th round, respectively, so that $W^{err}=\sum_i W^{err}_i$ and $W^{ec}=\sum_i W^{ec}_i$.  Let $N^{err}_i$ denote the number of errors (either spin flip or stabilizer) on the $i$-th round so that $N_{err}=\sum_i N^{err}_i$.
Note that if error correction produces spin flips in a ball and if the syndromes are correctly measured in that ball, it reduces the weight of the error chain by at least one; further, the only way to increase is to have a spin flip or stabilizer error measurement on a previous time slice.  Hence, $w_{i+1}\leq w_i - \ct W^{ec}_i+\ct N^{err}_i$ so
$W^{ec}_i \leq w_i-w_{i+1}+\ct N^{err}_i$.
So, $W^{ec}=\sum_{i=0}^{\tau-1} W^{ec}_i \leq -w_\tau+\ct N^{err}$ so
\be
W^{ec} \leq \ct N^{err}.
\ee 
So, $N_{err}\geq \ct W^{err}$ and $N_{err} \geq \ct W^{ec}$ so $N_{err} \geq \ct (W^{err}+W^{ec}) \geq \ct |E_{spacelike}|$, as claimed.

Let $|E_{timelike}|=W^g+W^p$,where $W^g$ is the sum of the weights of the chain before rounds without a poor choice and $W^p=\sum_{i \in \pc} w_{i}$ is the weight of the chain before rounds with a poor choice.  Note that the probability of having such a set of poor choices is bounded by $\exp(-\ct W^p)$.
We claim that $W^g \leq \ct N^{err}$.  To show this, note that if the $i$th round of error correction does not have a poor choice then $W^{ec}_i \geq \ct w_i$; here we use the analysis of the previous section which shows that without errors in the $i$-th round, we have $w_{i+1} \leq \cred w_i$ for $\cred<1$, and the fact that reducing the weight of the error chain by an amount $(1-\cred) w_i$ requires at least $\ct (1-\cred) w_i$ spin flips.
Hence, $W^g \leq \ct W^{ec}$.  However, as shown $W^{ec} \leq \ct N^{err}$.

Hence, the number of errors occuring within distance $\rdec$ of $E$ is lower bounded by $\ct(|E_{spacelike}|+W^g)$, and there are at most $O(1)^{|E|}$ possible ways to have such errors, while the probability of having the given set of poor choices is upper bounded by $\exp(-\ct W^p)$.  Hence, summing over possible sets of errors, the probability of having a set of errors and poor choices which gives a chain $E$ is bounded by
\be
P(E) \leq (\ct p)^{\ct |E_{spacelike}|}\cdot O(1)^{E_{timelike}},
\ee
as claimed.

We now turn to 
bounding the number of
$E$ with given $|E_{spacelike}|$ and $|E_{timelike}|$.   We begin with a simpler problem: 
bounding the number of $E$ containing $x$ with given $|E|$.  We show that there are $O(d)^{|E|}$ such $E$.
The set $E$ is connected by definition using the graph $G$ defined above.  This graph has degree $d=O(1)$, since we take $\rdec=O(1)$.  The set $E$ induces some connected subgraph of the graph; removing if necessary some edges from this subgraph, we obtain a tree $T$. There may be more than one way to remove edges from the graph to obtain a tree; we will count the number of trees which will give an upper bound to the number of subgraphs.  Now is the step where fixing $x$ is important: by fixing $E$ to contain $x$, we can take $x$ to be the root of the tree $T$.  We label the edges leaving each vertex in $G$ by integers $1,...,d$; if we remove the vertex labels from all nodes of $T$ other than the root, but still label the edges by these integers, then $T$ is still uniquely specified.  Thus, the tree $T$ is uniquely specified by specifying $x$ and specifying a rooted tree graph with the edges of the tree graph labelled by integers $1,...,d$, where all the integers connecting a given vertex to its daughters have distinct labels.  Equivalently, $T$ is uniquely specified by giving $x$ and giving a subtree of an infinite tree of degree $d$, with the subtree containing the root.  Such subtrees can be counted and there
are $O(d)^{|E|}$ such trees with $|E|$ vertices.

We now
bound the number of
$E$ with given $|E_{spacelike}|$ and $|E_{timelike}|$.
To deal with this different weighting of the spacelike and timelike cells, we ``coarse-grain" in time.  To do this,
let $k$ be an integer to be chosen below.  Let $\tilde w^{tot}(a)=\sum_i w_{ik+a}$ for integer $a=0,...,k-1$.  There must be some $a$ such that $k\tilde w^{tot}(a) \leq \sum_i w_i$.
Define a new graph $\tilde G$; this has the same vertex set as $G$.  The graph $\tilde G$ has an edge between vertices corresponding to two spacelike $2$-cells $q \times l$ and $r \times m$ if $l=ik+a+b$ and $m=ik+a+c$ for some integers $i$ and $0 \leq b,c \leq k-1$ and if $q$ and $r$ are within distance $2\rdec$ of each other.  The graph $\tilde G$ has an
edge between vertices corresponding to a spacelike $2$-cell $q \times l$ and a timelike $2$-cell $s \times m$ if $q$ and $s$ are attached to each other and $m=ik+a$ and
$l=ik+a+b$ or $l=(i-1)k+a+b$ for $0 \leq b \leq k-1$.  There is an edge between vertices corresponding to two timelike $2$-cells $r \times l$ and $s \times m$ if
$r=s$ and $l=ik+a$ and $m=(i+1)k+a$ or $l=(i+1)k+a$ and $m=ik+a$.  Note that the only timelike $2$-cells $s \times m$ connected to other $2$-cells are those with $m=ik+a$.
This graph $\tilde G$ in a sense ``rescales" the time direction by a factor of $k$.
Thus the degree of $\tilde G$ is $O(k^2)$.  Given a chain $E$, we define a connected subset of $\tilde G$ consisting of all vertices in $\tilde G$ corresponding to a spacelike $2$-cell in $E$ and also all vertices corresponding to a timelike $2$-cell $s \times m$ in $E$ with $m=ik+a$.
The number of such subsets with $v$ vertices is bounded by $O(k^2)^{v}$ since the degree of $\tilde G$ is $O(k^2)$.  Let $v_{timelike}$ be the number of vertices in this subset which correspond to a timelike $2$-cell and let $v_{spacelike}=v-v_{timelike}$.
We have that
\begin{eqnarray}
P(E)  & \leq & (\ct p)^{\ct |E_{spacelike}|}\cdot O(1)^{|E_{timelike}|} \\ \nonumber & \leq & (\ct p)^{\ct v_{spacelike}}\cdot O(1)^{k v_{timelike}}.
\end{eqnarray}
Choosing $k$ of order $\log(p)$, this is bounded by $(\ct p)^{v}$.  Hence, multiplying this by $O(k^2)^v$ to count the number of $E$ with given $v$, we find that
\be
\sum_E P(E) \leq \sum_{v \geq |E|} (\ct \log^2(p) p)^v.
\ee
Since $v \geq \ct \rinj$, Eq.~(\ref{ebnd}) follows.

\subsection{Error Correction to a Codeword}
\label{errcor}
 Finally, we consider the problem of error correcting back to a code word.  Suppose after some time $\tau$, the noise is turned off, both spin flip errors and stabilizer measurements.
Then, no connected components $C_a$ of the error chain $C$ can have have all their support after time $\tau$.  However, for any given component of the error chain with support before time $\tau$, the probability that it survives to a time $\tau+\delta$ is bounded by $((cp)^{\delta/c'}$.  Hence, after a time $\tau+\delta$ with $\delta \sim \log(N)$, the probability that any error chain survives is negligible.  So, in a logarithmic number of rounds (that is, almost constant parallelized time and hence almost linear total time) we succeed in correcting back to a codeword.  The correction process produces no errors in the codeword so long as all error chains have radius smaller than $\rinj$.  For $\tau$ which is at most polynomial in $N$, this occurs with high probability for sufficiently small $p$.

{\it Acknowledgments---} I thank M. Freedman and Z. Wang for useful discussions.

\end{document}